\documentclass[amsmath, amssymb, aps, prb, twocolumn, notitlepage]{revtex4-2}
\usepackage[utf8]{inputenc}
\usepackage{graphicx}               
\usepackage[caption=false]{subfig}  
\usepackage{dcolumn}                
\usepackage{bm}                     
\usepackage[dvipsnames]{xcolor}     
\usepackage{slashed}                
\usepackage{hyperref}               
\usepackage{enumitem}               
\usepackage{mathtools}              
\usepackage{floatrow}

\newcounter{myparagraphs}

\begin{document}
\title{Current Noise of Hydrodynamic Electrons}
\author{Aaron Hui}
\affiliation{Department of Physics, Ohio State University, Columbus, Ohio 43202, USA}
\author{Brian Skinner}
\affiliation{Department of Physics, Ohio State University, Columbus, Ohio 43202, USA}
\date{\today}

\begin{abstract}
    A resistor at finite temperature produces white noise fluctuations of the current called Johnson-Nyquist noise. Measuring the amplitude of this noise provides a powerful primary thermometry technique to access the electron temperature. In practical situations, however, one needs to generalize the Johnson-Nyquist theorem to handle spatially inhomogeneous temperature profiles. Recent work provided such a generalization for ohmic devices obeying the Wiedemann-Franz law, but there is a need to provide a similar generalization for hydrodynamic electron systems, since hydrodynamic electrons provide unusual sensitivity for Johnson noise thermometry but they do not admit a local conductivity nor obey the Wiedemann-Franz law. Here we address this need by considering low-frequency Johnson noise in the hydrodynamic setting for a rectangular geometry. Unlike in the ohmic setting, we find that the Johnson noise is geometry-dependent due to non-local viscous gradients. Nonetheless, ignoring the geometric correction only leads to an error of at most 40\% as compared to naively using the ohmic result.
\end{abstract}

\maketitle

\emph{Introduction} -- 
The measurement of heat flow has long been a pivotal tool for exploring many-body systems in condensed matter and materials physics.
For example, measurements of thermal conductivity or heat capacity reflect the existence of whichever quasiparticles are present in the system, including those which are charge neutral.
But such measurements require an accurate thermometer, and modern metrology schemes increasingly require nanoscale temperature resolution.

In the context of electron systems, one challenge of nanoscale thermometry is to disentangle electron and phonon contributions to heat transport; in settings with weak electron-phonon coupling, the electron and phonon temperatures may not even be the same. Johnson noise thermometry addresses the electronic half of this issue: it is a powerful primary thermometry technique that allows one direct and isolated access to the electronic degrees of freedom. For this technique's simplest formulation, consider a resistor held at a uniform electronic temperature $T_0$. The Johnson-Nyquist theorem (fluctuation-dissipation theorem) dictates that in a two-terminal setup
\begin{align}
    S(t-t') \equiv \langle \delta  I(t)  \delta I(t') \rangle = \frac{2k_B T_0}{R} \delta(t-t'),
    \label{eq: fluctuation dissipation}
\end{align}
where $\delta I(t) = I(t) - \langle I \rangle$ is the charge current fluctuation at time $t$ and $R$ is the resistance; $\langle \ldots \rangle$ denotes an ensemble average (i.e.\ a time average if ergodicity is assumed) \footnote{The usual expression $\langle I^2 \rangle = 4k_B T_0 \Delta f/R$ can be obtained by convolving with the inverse Fourier transform of a rectangular function (band-pass filter) of width $\Delta f$ (with low and high band edges set appropriately) and setting $t-t' = 0$. Since we are working with two-sided Fourier transforms, the negative frequencies will provide the additional factor of $2$.}. As seen in Eq.~\eqref{eq: fluctuation dissipation}, the time-averaged current fluctuations provide a direct measure of the temperature of the electron bath in the resistor. The noise correlator $S(t-t')$ can be written in units of temperature, defining the so-called Johnson noise temperature
\begin{align}
    T_\text{JN} \equiv \lim_{\omega\rightarrow 0} \frac{R}{2k_B} S(\omega)
\end{align}
where $S(\omega) = \int_{-\infty}^\infty dt \, e^{-i\omega t} S(t)$ is the (two-sided) Fourier transform of the current noise correlator. Therefore, in situations with uniform temperature the Johnson-Nyquist theorem tells us that $T_\text{JN} = T_0$; the Johnson noise temperature directly measures the electronic temperature $T_0$ without need for calibration. In other words, Johnson noise acts as a primary thermometer.
This thermometry technique has recently been fruitfully utilized to make record-sensitive bolometers \cite{Karasik2014, Efetov2018, Miao2018, Liu2018, Miao2021} and to make measurements of electronic thermal conductivity and heat capacity \cite{Fong2012,Crossno2015,Waissman2022}.

In many practical situations, such as those listed above, the fundamental Eq.~\eqref{eq: fluctuation dissipation} does not apply since the electronic temperature is not spatially uniform.  Generalizations of Eq.~\eqref{eq: fluctuation dissipation} were previously studied \cite{Sukhorukov1999,Pozderac2021} for electronic systems that obey Ohm's law, i.e.\ where a local proportionality $J(\mathbf{x}) = \sigma(\mathbf{x}) E(\mathbf{x})$ between current density and electric field holds, as well as the Wiedemann-Franz (WF) law. 
These studies find that when current flows through a two-terminal device, Joule heating leads to an increase in the measured Johnson noise (in temperature units) by
\begin{align}
    \delta T_\text{JN} \equiv T_\text{JN} - T_0 =  \frac{PR}{12 L_0 T_0}.
    \label{eq: ohmic Johnson Noise}
\end{align}
Here $P = I^2 R$ is the Joule power, $R$ is the resistance, and $L_0 = \kappa/(\sigma T_0) = (\pi^2/3) (k_B/e)^2$ is the Lorenz ratio. 
The quantity $\delta T_\text{JN}$ can be thought of as the excess noise arising from Joule heating; throughout this paper we refer to the total $T_\textrm{JN}$ as simply the ``Johnson noise". 
Equation (\ref{eq: ohmic Johnson Noise}) has been known for the special case of a rectangular geometry with a spatially uniform and diagonal conductivity tensor since at least 1992 \cite{Prober1992}, but in fact Eq.~(\ref{eq: ohmic Johnson Noise}) is generic for any two-terminal geometry and any form of the conductivity tensor (even if it exhibits spatial variations), so long as Ohm's law and the WF law are obeyed \cite{Pozderac2021}.

What has remained unknown is how Eq.~\eqref{eq: ohmic Johnson Noise} should be generalized for situations not governed by Ohm's law.
Electronic systems that violate Ohm's law have become increasingly prominent in recent years, with experimental works demonstrating a hydrodynamic regime of strongly-interacting electrons in a number of materials \cite{Lucas2018, Narozhny2022, deJong1995, Muller2009, Torre2015, Levitov2016, Bandurin2016, Crossno2016, Guo2017, Kumar2017, Bandurin2018, Braem2018, Bergdyugin2019, Gallagher2019, Sulpizio2019, Jenkins2020, Ku2020, Vool2021, Aharon-Steinberg2022, Moll2016, Bachmann2022, Gooth2018, Gusev2018, Levin2018, Gusev2020, Shavit2019, Stern2021, Kumar2022}.
In such systems there is no concept of a spatially local conductivity and WF is violated.
Graphene in particular is a material of choice both for Johnson noise thermometry and electron hydrodynamics, due to its relatively weak electron-phonon coupling, low disorder, and strong electron-electron interactions.
The advent of these hydrodynamic electron systems calls for an extension of Johnson noise theory to this new setting.

Moreover, a na\"ive application of Eq.~\eqref{eq: ohmic Johnson Noise} (as was used, for example, in the seminal measurements of WF violation in graphene \cite{Crossno2016}) suggests great practical utility of hydrodynamic electrons. 
Electrons in the hydrodynamic regime can display large WF violations \cite{Muller2008, Foster2009, Principi2015, Crossno2016, Lucas2016, Gooth2018, Lucas2018a, Zarenia2019, Zarenia2020, Robinson2021, Ahn2022, Li2022}, and deep in the hydrodynamic regime (with only a single type of carrier) the Lorenz ratio $\kappa/(\sigma T_0)$ becomes very small \cite{Muller2008, Foster2009, Principi2015, Lucas2018a, Ahn2022}.
Naively inserting this small effective Lorentz ratio into Eq.~\eqref{eq: ohmic Johnson Noise} suggests a very large sensitivity for Johnson noise in the hydrodynamic regime. 
Such high sensitivity would imply that hydrodynamic electrons are ideal for bolometry and thermometry applications.
Therefore, the key question of the fate of thermal noise in a hydrodynamic electron system and the validity of Eq.~\eqref{eq: ohmic Johnson Noise} has significant practical and theoretical importance for the development of electron thermometry.

\begin{figure}
    \centering
    \includegraphics[width=.9\columnwidth]{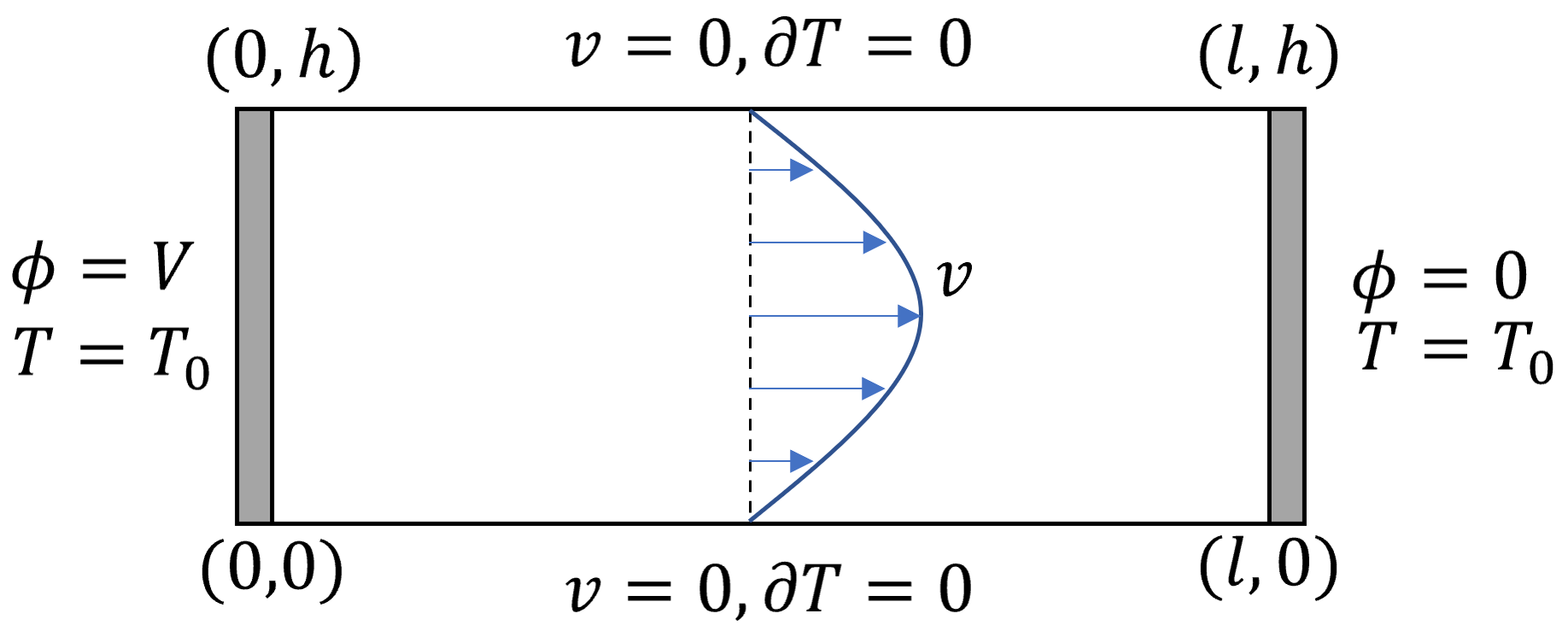}
    \caption{The rectangular geometry that we consider and its boundary conditions. A voltage $V$ is applied across the contacts, and we consider no-slip boundary conditions at the walls. We also fix the temperature on the $x$ boundaries and enforce $\partial_y T = 0$ on the $y$ boundaries.}
    \label{fig: BCs}
\end{figure}

In this paper, we explicitly study the Johnson noise of hydrodynamic electrons. 
We analytically solve for the low-frequency fluctuations of the Navier-Stokes equations in a rectangular geometry, depicted in Fig.~\ref{fig: BCs}. 
We find that the Johnson noise temperature is no longer geometry-independent due to non-local viscous gradients, as opposed to the ohmic case \cite{Pozderac2021}. 
Despite this nonuniversality, Eq.~\eqref{eq: ohmic Johnson Noise} is of the correct functional form up to a multiplicative geometric correction. 
In fact, this geometric correction is no larger than $40\%$ for any aspect ratio of the system or any value of the electron-electron scattering rate. 
Thus, Eq.~(\ref{eq: ohmic Johnson Noise}) provides a generally correct description of the Johnson noise, even though the resistance $R$ and Lorenz ratio $L$ are strongly renormalized by hydrodynamic effects.

\emph{Mathematical Setup} -- Throughout this paper, we work with the rectangular geometry shown in Fig.~\ref{fig: BCs}.
The full equations of motion for incompressible flow are given by
\begin{align}
    \partial_t \mathbf{v} + \mathbf{v} \cdot \nabla \mathbf{v} & = -\frac{1}{\rho} \nabla P - \frac{q}{m} \nabla \phi - \gamma\mathbf{v} + \nu \nabla^2 \mathbf{v}
    \label{eq: NS full}
    \\
    \partial_t T + \mathbf{v} \cdot \nabla T & = \frac{\kappa}{\rho c_p} \nabla^2 T + \frac{\nu}{2 c_p} \left(\partial_k v^i + \partial_i v^k\right)^2 + \frac{\gamma}{c_p} v^2
    \label{eq: heat full}
    \\
    \nabla \cdot \mathbf{v} & = 0.
    \label{eq: incompressibility full}
\end{align}
The hydrodynamic fields are the velocity $v$, the temperature $T$, the pressure $P$, and the electric potential $\phi$. 
The phenomenological constants in the equations of motion are the hydrodynamic mass $m$, the charge $q$, the mass density $\rho$, the momentum relaxation rate $\gamma$, the viscosity $\nu$, the specific heat at constant pressure $c_p$, and the thermal conductivity $\kappa$. 
We have assumed an incompressible flow with constant density \footnote{Incompressibility of flow has a \emph{different} meaning than that of the incompressibility of an electronic system. The former means only that $v \ll c$, while the latter is a statement about the thermodynamic compressibility $dn/d\mu$.}; this assumption is valid for flows with $v \ll c$ and $\tau \gg L/c$ where $c$ is the speed of sound and $\tau$ and $L$ are a characteristic time and length, respectively \cite{landauv6} \footnote{For an unscreened Coloumb potential in 3D, obeying the constitutive relation $\nabla^2 \delta \phi = \frac{\delta \rho_e}{\epsilon}$, the incompressibility assumption is valid for $v \ll \omega_p L$, where $\omega_p = \sqrt{\frac{ne^2}{m \epsilon}}$ is the plasma frequency for $n$ the number density and $\epsilon$ the electric permittivity. Similarly, for the unscreened 2D Coloumb potential, where $\omega \sim \sqrt{k}$, one should take $c$ to be the smallest characteristic group velocity, i.e. $c\sim \left.\frac{d\omega}{dk}\right|_{k=1/L}$.}. 
We also neglect the pressure $P$ since it can be subsumed into an effective electric potential $\phi' = \phi + mP/(\rho q)$. 
Finally, we will work at linear order, neglecting convection terms $\mathbf{v}\cdot\nabla \mathbf{v}$ and thermal advection $\mathbf{v}\cdot \nabla T$. 
Dropping convection is valid at low Reynolds numbers $\operatorname{Re}_\nu \equiv vL/\nu \ll 1$ \footnote{Here we invoke Reynolds numbers only to justify dropping convection terms from the hydrodynamic equations, and not to estimate the conditions for the onset of turbulence, which in general is suppressed by the presence of static disorder (as measured by the momentum-relaxing Reynolds number).} or at low ``momentum-relaxation Reynolds number''\cite{Hui2021} $\operatorname{Re}_\gamma \equiv v/L\gamma \ll 1$. 
Graphene experiments are typically deep within this low Reynolds number regime, with $\gamma \sim 650$ GHz, $\nu \sim 0.1 \text{m}^2/\text{s}$, and $L \sim 5\, \mu$m \cite{Bandurin2016} so that $\operatorname{Re}_\gamma \sim I/(26 \text{mA})$ and $\operatorname{Re}_\nu \sim I/(160 \mu\text{A})$ \cite{Hui2021}.
Moreover, dropping thermal advection is valid for $\kappa/(c_p L) \gg v$ when thermal diffusion is fast compared to the fluid velocity. After these simplifications, the equations of motion become
\begin{align}
    \partial_t \mathbf{v} & = - \frac{q}{m} \nabla \phi - (\gamma - \nu \nabla^2) \mathbf{v}
    \label{eq: NS}
    \\
    \partial_t T & = \frac{\kappa}{\rho c_p} \nabla^2 T + \frac{\nu}{2 c_p} \left(\partial_k v^i + \partial_i v^k\right)^2 + \frac{\gamma}{c_p} v^2
    \label{eq: heat}
    \\
    \nabla \cdot \mathbf{v} & = 0
    \label{eq: incompressibility}
\end{align}
Eq.~\eqref{eq: NS} is the momentum balance equation with an electric force, momentum relaxation, and viscous drag. Eq.~\eqref{eq: heat} is the heat equation, with source terms from viscous heating and Joule heating. 
Supplemented by fixed-voltage, fixed-temperature, and no-slip/no-heat-flow boundary conditions (see Fig.~\ref{fig: BCs}), solving these equations provides the quasi-equilibrium solution about which we will study noise fluctuations.

Once the steady-state solution is known, we can study the thermal noise fluctuations. We are interested in low frequency solutions, $s \rightarrow 0$, where the velocity fluctuations $\delta v_i$ are described by the Laplace transform of Eq.~\eqref{eq: NS} \cite{landauv9}
\begin{align}
    (s +\gamma - \nu \nabla_\mathbf{r}^2 ) \langle \delta v_i(\mathbf{r},s) \delta v_j(\mathbf{r}',0)\rangle & = \langle \delta v_i(\mathbf{r},0) \delta v_j(\mathbf{r}',0)\rangle 
    \nonumber 
    \\
    & \equiv \frac{k_B T(\mathbf{r})}{\rho} \delta(\mathbf{r}-\mathbf{r'}) \delta_{ij}
    \label{eq: Johnson noise}
\end{align}
where $s$ is the Laplace parameter \footnote{Since the correlation function is time-reversal even, from the Laplace transform we can obtain the Fourier transform via the replacement $s\rightarrow i\omega$ \cite{landauv9}.} and the initial condition on the RHS of Eq.~\eqref{eq: Johnson noise} is given by the equipartition theorem. This equation is supplemented by incompressibility of fluctuations $\partial_i \langle \delta v_i(\mathbf{r},s) \delta v_j(\mathbf{r}',0)\rangle = 0$. In writing Eq.~\eqref{eq: Johnson noise} with incompressibility, we have neglected the electric potential (pressure) and density fluctuations. This approximation is again valid when $\omega \ll c/L$, i.e.\ when the frequency is much smaller than the characteristic sound frequency of the sample. Finally, to obtain the current fluctuations from the velocity fluctuations, it is convenient to apply the relation $I = \frac{1}{\ell}\int dx dy J_x$ to the solution of Eq.~\eqref{eq: Johnson noise}; this relation arises from current conservation since $\int dy J_x$ is independent of $x$. 

For simplicity, we ignore other forms of noise that may be present. Shot noise, in particular, is also present; for diffusive conductors it dominates whenever the source-drain voltage $V$ is large enough that $eV \gg k_B T$ (see, e.g., Refs.~\cite{kogan_electronic_1996, Nagaev1995}). 
However, recent experiments measuring Johnson noise in hydrodynamic electrons tend to operate in a regime where Johnson noise is dominant over shot noise \cite{Fong2012, Crossno2015, Crossno2016, Waissman2022} \footnote{The dominance of Johnson noise is experimentally evidenced by the linear relationship between heating power and noise power. This could be because hydrodynamic behavior tends to appear at elevated temperatures (${\gtrsim 100}$ K), but one may also expect electron-electron interactions to suppress shot noise due to strong positional correlations between electrons.}. 
We leave an exploration of other noise mechanisms in the hydrodynamic regime to future work.

\emph{Solution} -- We begin by solving the steady-state equations of motion to determine the temperature profile. 
The steady-state velocity profile is given by the ohmic-Poiseuille solution (see, e.g., Ref.~\onlinecite{Torre2015})
\begin{align}
    J_x(y) \equiv n v_x(y) = \sigma_D E_x \left[1 - \frac{\cosh \left( \frac{y-h/2}{\lambda} \right) }{\cosh \left( \frac{h}{2\lambda} \right) }\right],
    \label{eq: current profile}
\end{align}
where $\sigma_D \equiv ne^2/(m\gamma)$ is the Drude conductivity and the electric field $E_x = V/\ell$. The viscous length scale or Gurzhi length $\lambda \equiv \sqrt{\nu/\gamma}$ is a length scale below which viscous effects are important.
For convenience, we can define an effective conductivity $\overline{\sigma}\equiv (\ell/h) / R = \left(1-\frac{2\lambda}{h}\tanh\frac{h}{2\lambda}\right)\sigma_D$, where the two-terminal resistance $R = V/I$ was computed using Eq.~\eqref{eq: current profile}. 
In the ohmic limit $\lambda \ll h$, the effective conductivity $\overline{\sigma} \rightarrow \sigma_D$ reduces to the usual Drude conductivity. 
We emphasize that a local conductivity is \emph{not} well-defined in the presence of viscosity; the effective conductivity is useful as a measure of the heat dissipation, not of the local current-voltage relation. 
The solution for the velocity profile in Eq.~\eqref{eq: current profile} determines the dissipative heating terms in Eq.~\eqref{eq: heat}, which allows us to solve for the temperature profile. 

\begin{figure}
    \centering
    \begin{sidesubfloat}[][]{
    \includegraphics[width=.75\columnwidth]{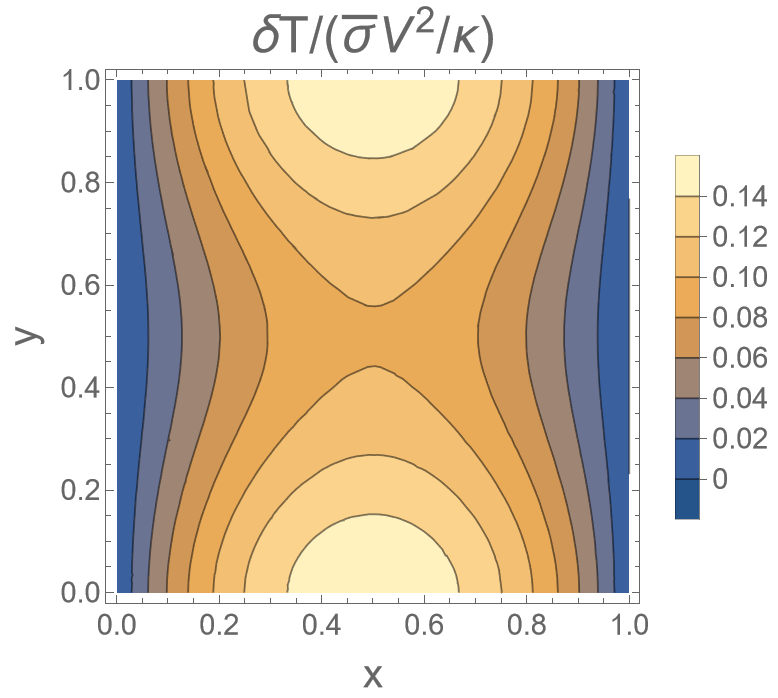}
    \label{fig: viscous temp profile original}
    }
    \end{sidesubfloat}
    
    \vspace{.5em}
    
    \begin{sidesubfloat}[][]{
    \includegraphics[width=.75\columnwidth]{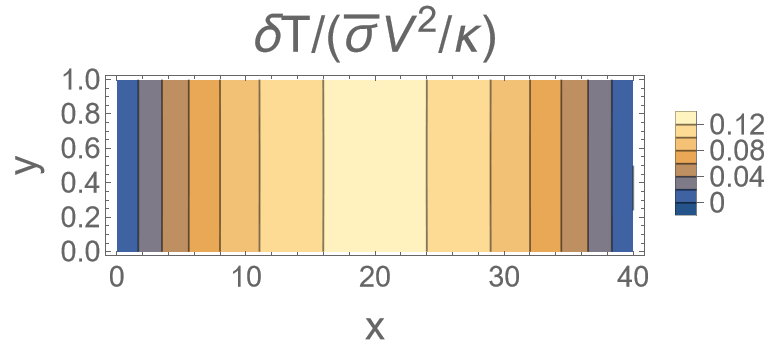}
    \label{fig: viscous temp profile thin}
    }
    \end{sidesubfloat}
    \caption{A plot of the purely viscous ($\lambda\rightarrow \infty$) temperature profiles for (a) $h/\ell=1$ and (b) $h/\ell=1/40$. Notice that for (b), the profile is very similar to the ohmic profile, having negligible $y$ variation.
    }
    \label{fig: viscous temp profiles}
\end{figure}

We use Fourier techniques to obtain the temperature profile analytically.
The exact result can be written as 
\begin{align}
    T(\mathbf{x}) = T_0 + \frac{\overline{\sigma} V^2}{\kappa}\left[\frac{1}{2}\tilde{x}(1-\tilde{x}) - \frac{h^2}{\ell^2}F\left(\frac{h}{\ell}, \frac{\lambda}{h}; \mathbf{x}\right)\right]
    \label{eq: temperature result}
\end{align}
where $\tilde{x} = x/l$ is the non-dimensionalized $x$-coordinate and $F$ is a double Fourier sum \footnote{See Supplemental Material [url] for mathematical details, which includes Refs. \cite{Conca1994, landauv9, Narozhny2019}}. 
To obtain a qualitative understanding, we consider this result in various simplifying limits (see Fig.~\ref{fig: viscous temp profiles}).

In the ohmic limit $\lambda \ll h$, the heating profile (i.e. the heat per unit area generated at each point) is spatially uniform; since $F\rightarrow 0$, this limit admits the simple parabolic temperature profile
\begin{align}
    T_\text{ohm}(x) = T_0 + \frac{\overline{\sigma} V^2}{\kappa} \frac{1}{2}\tilde{x}(1-\tilde{x}) + \mathcal{O}(\lambda/h),
\end{align}
with $\overline{\sigma} \rightarrow \sigma_D$. This parabolic profile is the result of the thermal boundary conditions: heat is only allowed to flow at the contacts at $x=0$ and $x=\ell$, so the temperature must be maximal in the center and minimal at the fixed-temperature boundaries.

In the opposite viscous limit, $\lambda \gg h$, viscous dissipation leads to a non-uniform heating along the $y$ direction (though heating is still uniform in $x$, as in the ohmic case). The temperature profile then obtains a ``geometric correction'' to the ohmic result as a function of the aspect ratio $h/\ell$. 
In the thin channel regime $h \ll \ell$, analyticity of $F$ around $h/\ell = 0$ means that the temperature profile takes the simple form
\begin{align}
    T_\text{thin} = T_0 + \frac{\overline{\sigma} V^2}{\kappa}\frac{1}{2}\tilde{x}(1-\tilde{x}) + \mathcal{O}(h^2/\ell^2)
\end{align}
This ``ohmic-like'' temperature profile arises due to the fact that in the thin channel limit, the problem becomes quasi-1D; heating inhomogeneities rapidly equilibrate along $y$ as compared to along $x$, leading to small, $\mathcal{O}(h^2/\ell^2)$ corrections to the ``ohmic-like'' temperature profile that is independent of $y$. 
An example of an ``ohmic-like" temperature profile in the thin channel limit is plotted in Fig.~\ref{fig: viscous temp profile thin}.

As we will see below, the temperature profile determines the Johnson noise profile, so that correspondingly Eq.~\eqref{eq: ohmic Johnson Noise} is obeyed in both the ohmic limit and the thin channel limit (for the latter, even in the limit of zero Drude resistivity, $\gamma \rightarrow 0$).
Therefore, in order to observe any deviation from Eq.~\eqref{eq: ohmic Johnson Noise}, one needs to be in the regime $\ell \gtrsim h \gtrsim \lambda$, for which the temperature profile is nonuniform in the $y$ direction.  
An example is shown in Fig.~\ref{fig: viscous temp profile original}, where one can see increased temperature at the boundaries due to viscous dissipation from the no-slip boundary friction.

\floatsetup[figure]{style=plain,subcapbesideposition=top}

Given the temperature profile, we can solve for the measured current noise via Eq.~\eqref{eq: Johnson noise}, again using Fourier techniques. 
The analytic result can be written as
\begin{align}
    S = \frac{2k_B}{R}\left[T_0 + \frac{\overline{\sigma} V^2}{12\kappa}f(h/\ell,\lambda/h)\right],
\end{align}
where the function $f$ is a Fourier series and is plotted in Fig.~\ref{fig: geometric correction} \footnote{See the Supplemental Material [url] for mathematical details.}.
In the ohmic limit $\lambda \ll h$, we find that $f \rightarrow 1$, recovering the previous ohmic result for the Johnson noise \cite{Pozderac2021} (see Eq.~\eqref{eq: ohmic Johnson Noise}).
\begin{align}
    S_\text{ohm} = \frac{2k_B}{R}\left[T_0 + \frac{\sigma_D V^2}{12\kappa}\right]
\end{align}
Therefore, we interpret $f$ as a geometric correction to the ohmic Johnson noise result.
For $h \ll \ell$ where the temperature profile is ``ohmic-like'', we also obtain the ohmic result $f \rightarrow 1$. Around $\lambda/h \sim 0.2$, we find that $f-1$ changes sign. 
This sign change is due to a crossover from the regime where Joule heating dominates to the regime where viscous heating dominates, which produces a corresponding change in the ``topography'' of the temperature profile. 
When Joule heating dominates ($\lambda/h \lesssim 0.2$) there is a single temperature peak in the center, while when viscous heating dominates ($\lambda/h \gtrsim 0.2$) there are two temperature peaks, one at each boundary. 

In general, the dimensionless function $f$ is a slow $\mathcal{O}(1)$ function of $h/\ell$ and $\lambda/h$ and $f$ never deviates from unity by more than $40\%$. 
Deep in the limit of viscous flow and large aspect ratio ($\lambda/h \gg 1$ and $h/\ell\gg 1$), i.e.\ the top-right corner of Fig.~\ref{fig: geometric correction}, the value of $f$ approaches $3/5$. 
Despite this conclusion that the ``ohmic'' result of Eq.~\eqref{eq: ohmic Johnson Noise} is always ``nearly correct,'' we emphasize that the sensitivity of Johnson noise thermometry is generally strongly renormalized by hydrodynamic effects. 
Specifically, viscous effects tend to strongly renormalize the resistance $R$ and the Lorenz ratio $L = \kappa/(\overline{\sigma} T_0)$, thereby making a large quantitative change in the measured Johnson noise.

\begin{figure}
    \centering
    \includegraphics[width=.95\columnwidth]{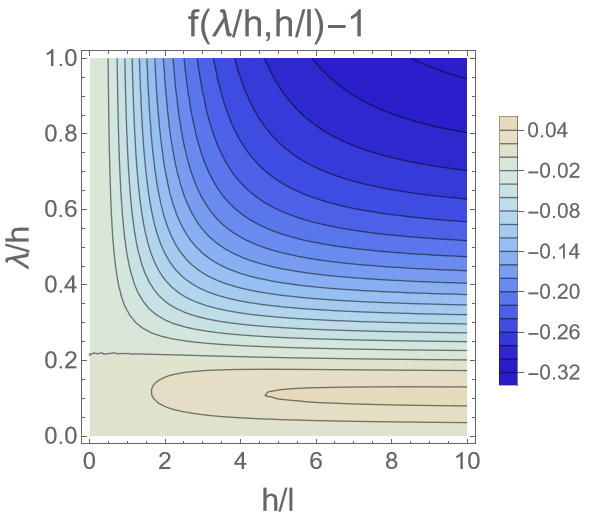}
    \caption{A plot of $f(h/l, \lambda/h)-1$, the deviation of the geometric correction to the Johnson noise from unity. In both the ohmic limit, $\lambda/h \ll 1$, and the thin channel limit, $h/l\ll 1$, we find $f\rightarrow 1$. Viscous effects are most prominent for $\lambda/h \gg 1$ and $h/l \gg 1$, where $f\rightarrow 3/5$.
    }
    \label{fig: geometric correction}
\end{figure}

\emph{Conclusion} --  In this paper, we have shown that the relationship between Johnson noise and heating for ohmic and WF-obeying systems [Eq.~\eqref{eq: ohmic Johnson Noise}] is, surprisingly, mostly valid even for hydrodynamic electrons [Eq.~\eqref{eq: Johnson noise}]. 
A geometric correction arises from preferential heating near the no-slip boundaries by viscous shear, but this correction is never more than $40\%$, regardless of the sample's aspect ratio or viscosity. Our result enables a range of fundamental and applied applications in thermometry and bolometry, and justifies applying existing Johnson noise thermometry techniques (those of Refs.~\onlinecite{Crossno2015} and \onlinecite{Waissman2022}, for example) directly in the hydrodynamic regime.

Our results, derived for a Galilean-invariant fluid, can be directly extended to the ``Dirac fluid'' limit where n-type and p-type carriers coexist (as in graphene near the charge neutral point) \footnote{Additional contributions due to relativistic hydrodynamics are negligible in the limit $v \ll v_F$, where $v_F$ is the ``speed of light.'' \cite{Lucas2018}}. 
In general, with chemical potential $\mu$ away from the Dirac point, electron-hole scattering causes the majority carriers to drag the minority carriers, so that electrons and holes equilibrate to the same hydrodynamic drift velocity. 
Very near the Dirac point, however, there is a zero-momentum mode with disequilibrated electron and hole drift velocities that can also carry current. 
This zero-momentum mode can relax current via electron-hole scattering, increasing the current-relaxation rate and suppressing the viscous length $\lambda$. 
Therefore, when the chemical potential is sufficiently close to the Dirac point, the current noise should return to ohmic-like behavior. 
To estimate the window where the zero-momentum mode is important, we estimate the two modes' relative contribution to the effective conductivity. We find \footnote{See Supplemental Material [url] for details of the estimate.}
\begin{align}
    \frac{\sigma_z}{\sigma_F} \sim \frac{(k_B T)^2}{\mu^2} \left[\frac{l_\text{ee}^2}{L^2} + \frac{\gamma_\text{imp}}{\gamma_\text{ee}}\right],
    \label{eq: dirac fluid estimate}
\end{align}
where $\sigma_z$ and $\sigma_F$ correspond to the zero-momentum and finite-momentum (hydrodynamic) conductivities, respectively, and $L$ is the sample length.
Therefore, even in the Dirac fluid limit, the zero-momentum mode can be neglected so long as $\mu^2/(k_B T)^2 \gg l_\text{ee}^2/L^2 + \gamma_\text{imp}/\gamma_\text{ee}$, where the RHS is small in the hydrodynamic limit. 
Where this inequality is satisfied, our main result, Eq.~\eqref{eq: Johnson noise}, applies directly; where it is violated, the ohmic result Eq.~\eqref{eq: ohmic Johnson Noise} applies. 
We remark that one can make more rigorous estimates using the explicit expressions from Boltzmann kinetic theory \cite{Gallagher2019, Sun2018, Muller2008} which give the same functional form.


While for a rectangular geometry the geometric correction $f$ to the Johnson noise does not deviate greatly from unity, one may wonder whether this conclusion is strongly geometry-dependent. More specifically, one can ask about the annular Corbino geometry; it exhibits ``paradoxical'' behavior for hydrodynamic electron flow, with a near-vanishing of the bulk electric field even when a strong current is flowing \cite{Shavit2019, Stern2021, Kumar2022, Hui2022, Levchenko2022}.
If one na\"ively applies the Shockley-Ramo theorem \cite{Shockley1938, Ramo1939, He01, Song2014}, then this bulk electric field expulsion would seemingly imply that the Johnson noise is unmodified by current flow, even as this flow produces significant electron heating. 
However, the version of Shockley-Ramo appropriate for an electron fluid \cite{Song2014} relies on a well-defined local conductivity. 
We expect that the current noise for the Corbino geometry is qualitatively similar to the rectangle; we do not expect any zeros or anomalies, only a quantitative change in the geometric correction. 
In the ohmic limit (i.e.\ small $\lambda$), we expect a return to Eq.~\eqref{eq: ohmic Johnson Noise}. 
Moreover, when the two annular radii $r_2 - r_1\equiv \delta \ll r_1$ are very close, the temperature variations are suppressed by $\mathcal{O}\left(\delta^2/r_1^2 \right)$ so we also expect the current noise to be ohmic-like in this 1D limit. 
We leave further exploration of the Corbino and of other geometries to future work.

\vspace{5mm}
{\noindent\bf Acknowledgements}
We thank Philip Kim, Jonah Waissman, Artem Talanov, Zhongying Yan, Alex Levchenko, Gregory Falkovich, Daniel E.\ Prober, and Justin C.\ W.\ Song for discussions. A.\ H.\ was partially supported by the Center for Emergent Materials, an NSF-funded MRSEC, under Grant No. DMR-2011876. B.\ S.\ was partly supported by NSF Grant No. DMR-2045742.

\bibliography{biblio}

\appendix
\onecolumngrid
\section{Calculational details}

In this appendix, we fill in the calculational details of the main text.
We first consider the heat equation Eq.~\eqref{eq: heat}. Upon computing the heating terms, the heat equation becomes
\begin{align}
    \nabla^2 T =&- \frac{\sigma_D E_x^2}{\kappa} \left[\left(\frac{\sinh \frac{y-h/2}{\lambda}}{\cosh\frac{h}{2\lambda}}\right)^2 + \left(1 - \frac{\cosh \frac{y-h/2}{\lambda}}{\cosh\frac{h}{2\lambda}} \right)^2\right]
    \label{eq: appendix heat}
\end{align}
where the first term on the RHS corresponds to viscous heating and the second term to Joule heating. To solve this, we look for a Fourier series solution of the form
\begin{align}
    T = T_0 + \sum_{a_x=1}^\infty \sum_{b_x=1}^\infty T_{a_x b_y} \sin(\pi a_x \tilde{x}) \cos(\pi b_y \tilde{y})
\end{align}
where $\tilde{x} = x/l$ and $\tilde{y} = y/h$ are the non-dimensionalized coordinates. This solution satisfies the thermal boundary conditions by construction. Therefore, we need to find the corresponding Fourier series of the heating terms on the RHS of Eq.~\eqref{eq: appendix heat}. This is given by
\begin{align}
    &\frac{\sigma_D E_x^2}{\kappa} \left[\left(\frac{\sinh \frac{y-h/2}{\lambda}}{\cosh\frac{h}{2\lambda}}\right)^2 + \left(1 - \frac{\cosh \frac{y-h/2}{\lambda}}{\cosh\frac{h}{2\lambda}} \right)^2\right] \nonumber
    \\
    =& \frac{\overline{\sigma} E_x^2}{\kappa}\left[\sum_{a_x=1}^\infty \frac{4\sin^2 \frac{\pi a_x}{2}}{\pi a_x}\sin(\pi a_x \tilde{x})\right]\left[1 - \sum_{b_y=1}^\infty \frac{1}{1-2\tilde{\lambda}\tanh \frac{1}{2\tilde{\lambda}}}\frac{4\tilde{\lambda}\tanh\frac{1}{2\tilde{\lambda}}\left(1-2\tilde{\lambda}^2\pi^2 b_y^2\right)}{\left(1 + \tilde{\lambda}^2\pi^2 b_y^2\right)\left(1+ 4\tilde{\lambda}^2\pi^2 b_y^2\right)} \cos(2\pi b_y \tilde{y})\right]
\end{align}
where the first bracketed term on the RHS corresponds to the Fourier series of unity and $\tilde{\lambda} \equiv \lambda/h$ is the non-dimensionalized viscous length. Therefore, the solution for the temperature profile is
\begin{align}
    T =& T_0 + \frac{\overline{\sigma}V^2}{\kappa} \Bigg[\frac{1}{2}\tilde{x}(1 -\tilde{x}) \nonumber
    \\
    &- \frac{h^2}{l^2}\sum_{a_x=1}^\infty\sum_{b_y=1}^\infty \frac{4\sin^2 \frac{\pi a_x}{2}}{\pi a_x} \frac{1}{\frac{h^2}{l^2}\pi^2 a_x^2 + 4\pi^2 b_y^2} \frac{1}{1-2\tilde{\lambda}\tanh \frac{1}{2\tilde{\lambda}}}\frac{4\tilde{\lambda}\tanh\frac{1}{2\tilde{\lambda}}\left(1-2\tilde{\lambda}^2\pi^2 b_y^2\right)}{\left(1 + \tilde{\lambda}^2\pi^2 b_y^2\right)\left(1+ 4\tilde{\lambda}^2\pi^2 b_y^2\right)} \sin(\pi a_x \tilde{x}) \cos(2\pi b_y \tilde{y})\Bigg]
    \\
    \equiv& T_0 + \frac{\overline{\sigma}V^2}{\kappa} \Bigg[\frac{1}{2}\tilde{x}(1 -\tilde{x}) -\frac{h^2}{l^2}F\left(\frac{h}{l},\frac{\lambda}{h}; \mathbf{x}\right)\Bigg]
\end{align}
whivh validates the form of Eq.~\eqref{eq: temperature result}.

Then, we solve Eq.~\eqref{eq: Johnson noise} for the Johnson noise. Similar to before, to satisfy the boundary conditions we take the following Fourier expansions
\begin{align}
    \langle \delta v_x (\mathbf{r},s) \delta v_x(\mathbf{r}', 0)\rangle =& \sum_{m_x=-\infty}^\infty \sum_{n_y=1}^\infty A_{m_x n_y} e^{2\pi i m_x \tilde{x}} \sin(\pi n_y \tilde{y})
    \\
    \frac{k_B T(\mathbf{r})}{\rho}\delta(\mathbf{r} - \mathbf{r}') =& \frac{k_B T(\mathbf{r}')}{\rho}\sum_{m_x=-\infty}^\infty \sum_{n_y=1}^\infty \frac{2}{hl}e^{2\pi i m_x (\tilde{x}-\tilde{x'})} \sin(\pi n_y \tilde{y}')\sin(\pi n_y \tilde{y})
\end{align}
In particular, our $\langle \delta v_x \delta v_x \rangle$ ansatz satisfies the no-slip boundary conditions. Therefore, this gives us the solution
\begin{align}
    \langle \delta v_x (\mathbf{r},s) \delta v_x(\mathbf{r}', 0)\rangle \equiv& \sum_{m_x=-\infty}^\infty \sum_{n_y=1}^\infty \frac{1}{s+\gamma+\nu \frac{\pi^2 n_y^2}{h^2}} \frac{k_B T(\mathbf{r}')}{\rho}\frac{2}{hl} e^{-2\pi i m_x (\tilde{x} - \tilde{x'})}\sin(\pi n_y y') \sin(\pi n_y y)
\end{align}
For (existence and) uniqueness statements, see Ref.~\onlinecite{Conca1994}.
Since the velocity-velocity correlator is time-reversal even, the Fourier transform of the velocity-velocity correlator is given by the replacement $s\rightarrow i\omega$ \cite{landauv9} with an additional overall factor of $2$.
In the $\omega\rightarrow 0$ limit and using $\delta\overline{I_x} = ne \int dx\,  dy\,  \delta v_x$, we get
\begin{align}
    \langle \delta \overline{I_x}(\omega\rightarrow 0) \delta \overline{I_x}(0)\rangle =&2k_B \frac{h}{l}\sigma_0 \sum_{n_y=1}^\infty \frac{1}{1+\tilde{\lambda}^2\pi^2 n_y^2}  \frac{4 \sin^2\frac{\pi n_y}{2}}{\pi n_y} \frac{1}{lh} \int_0^l dx' \int_0^h dy' T(\mathbf{r}') \sin(\pi n_y \tilde{y}')
    \\
    \equiv& \frac{2k_B}{R}\left[T_0 + \frac{\overline{\sigma} V^2}{12 \kappa}f(h/l, \lambda/h)\right]
\end{align}
The geometric correction $f$ is evaluated as
\begin{align}
    f(h/l,\lambda/h) \equiv& \sum_{n_y=1}^\infty \frac{1}{1-2\tilde{\lambda}\tanh\frac{1}{2\tilde{\lambda}}}\frac{1}{1+\tilde{\lambda}^2\pi^2 n_y^2}  \frac{4 \sin^2\frac{\pi n_y}{2}}{\pi n_y} \frac{1}{lh} \int_0^l dx' \int_0^h dy' \delta T(\mathbf{r}') \sin(\pi n_y \tilde{y}')
    \\
    =& 1 - \frac{12 h^2}{l^2}\sum_{b_y=1}^\infty\sum_{n_y=1}^\infty \sum_{a_x=1}^\infty \frac{4\tilde{\lambda}\tanh\frac{1}{2\tilde{\lambda}}}{\left(1-2\tilde{\lambda}\tanh\frac{1}{2\tilde{\lambda}}\right)^2}\frac{1}{1+\tilde{\lambda}^2\pi^2 n_y^2}  \frac{8\sin^4 \frac{\pi a_x}{2}}{\pi^2 a_x^2} \frac{1}{\frac{h^2}{l^2}\pi^2 a_x^2 + 4\pi^2 b_y^2} \nonumber
    \\
    & \phantom{1- \frac{12 h^2}{l^2}\sum_{b_y=1}^\infty\sum_{n_y=1}^\infty \sum_{a_x=1}^\infty}\times \frac{1-2\tilde{\lambda}^2\pi^2 b_y^2}{\left(1 + \tilde{\lambda}^2\pi^2 b_y^2\right)\left(1+ 4\tilde{\lambda}^2\pi^2 b_y^2\right)} \frac{8 \sin^4\frac{\pi n_y}{2}}{\pi^2 n_y^2 - 4 \pi^2 b_y^2}
    \\
    =& 1- \frac{12 h^2}{l^2}\sum_{b_y=1}^\infty\sum_{n_y=1}^\infty \frac{4\tilde{\lambda}\tanh\frac{1}{2\tilde{\lambda}}}{\left(1-2\tilde{\lambda}\tanh\frac{1}{2\tilde{\lambda}}\right)^2} \frac{1}{4\pi^2 b_y^2} \left(1 - \frac{h}{l\pi b_y} \tanh\frac{l\pi b_y}{h}\right) \frac{1-2\tilde{\lambda}^2\pi^2 b_y^2}{\left(1 + \tilde{\lambda}^2\pi^2 b_y^2\right)\left(1+ 4\tilde{\lambda}^2\pi^2 b_y^2\right)} \nonumber
    \\
    &\phantom{1- \frac{12 h^2}{l^2}\sum_{b_y=1}^\infty\sum_{n_y=1}^\infty }\times \frac{1}{1+\tilde{\lambda}^2\pi^2 n_y^2} \frac{8 \sin^4\frac{\pi n_y}{2}}{\pi^2 n_y^2 - 4 \pi^2 b_y^2}
    \\
    =&1 + \frac{6 h^2}{l^2}\sum_{b_y=1}^\infty \left(\frac{2\tilde{\lambda}\tanh\frac{1}{2\tilde{\lambda}}}{1-2\tilde{\lambda}\tanh\frac{1}{2\tilde{\lambda}}}\right)^2 \frac{1}{\pi^2 b_y^2} \frac{1-2\tilde{\lambda}^2\pi^2 b_y^2}{\left(1 + \tilde{\lambda}^2\pi^2 b_y^2\right)\left(1+ 4\tilde{\lambda}^2\pi^2 b_y^2\right)^2} \left(1 - \frac{h}{l\pi b_y} \tanh\frac{l\pi b_y}{h}\right)
    \label{eq: appendix full geom result}
\end{align}
From this expression, it is immediate that $f\rightarrow 1$ when $h/l \rightarrow 0$ or $\lambda/h \rightarrow 0$; the geomtric correction $f$ becomes unity and the Johnson noise temperature returns to the ohmic result for an ohmic (ohmic-like) temperature profile. 

\begin{figure}
    \centering
    \includegraphics[width=.4\columnwidth]{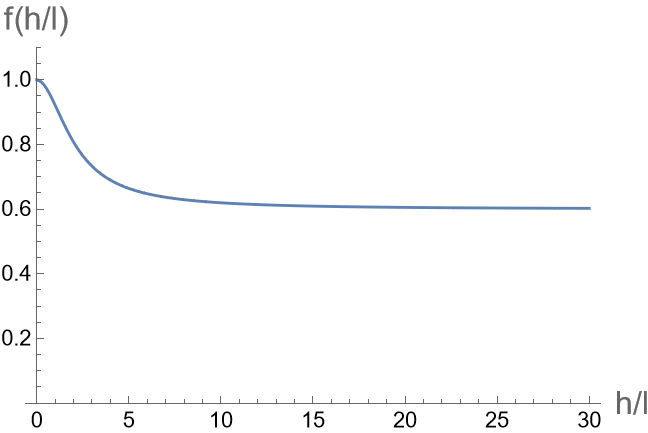}
    \caption{A plot of the geometric correction $f(h/l)$ to the effective Johnson noise temperature in the viscous limit $\lambda/h \rightarrow \infty$. It takes the limits $f(0) = 1$ and $f(h/l\rightarrow \infty) = 3/5$}
    \label{fig: appendix geom correction}
\end{figure}

In the viscous limit $\lambda/h \rightarrow \infty$, the geometric correction Eq.~\ref{eq: appendix full geom result} simplifies to
\begin{align}
    f(h/l) \equiv f(h/l,\lambda\rightarrow\infty) = 1 - \frac{h^2}{l^2}\sum_{b_y=1}^\infty \frac{108}{\pi^6 b_y^6}  \left(1 - \frac{h}{l\pi b_y} \tanh\frac{l\pi b_y}{h}\right)
\end{align}
This is plotted in Fig.~\ref{fig: appendix geom correction} and has the limits
\begin{align}
    f(h/l \ll 1) =& 1 - \frac{4}{35} \frac{h^2}{l^2}+\mathcal{O}\left(\frac{h^3}{l^3}\right)
    \\
    f(h/l \gg 1)=& 1-\frac{2}{5} +\mathcal{O}\left(\frac{l^2}{h^2}\right)
\end{align}

\section{Estimate for the zero-momentum mode of the Dirac fluid}
Here, we fill in the mathematical details of estimating when one can neglect the zero-momentum mode in the Dirac fluid [Eq.~\eqref{eq: dirac fluid estimate}].
Using the Drude form $ne^2/(m\gamma)$ as a guide, we need to estimate the parameters $n$, $m$, and $\gamma$ for the two modes. 
We first consider the finite-momentum (hydrodynamic) mode. The effective conductivity of this mode is $\sigma_F \sim \frac{n_- e^2}{W/(n_- v_F)^2} \frac{1}{\gamma_\text{imp} + \nu/L^2}$, where $W$ is the enthalpy, $n_- = n_e - n_h$ is the charge density, $v_F$ is the Dirac Fermi velocity, $\gamma_\text{imp}$ is the momentum-relaxing scattering rate, and $L$ is a characteristic length. 
For the zero-momentum mode, we estimate the effective conductivity to be $\sigma_z \sim \frac{n_+ e^2}{k_B T/v_F^2} \frac{1}{\gamma_\text{ee}}$, where $n_+ = n_e + n_h$ is the particle density, $\gamma_\text{ee}$ is the electron-electron collision rate, and we have estimated the mass with the thermal mass $k_B T/v_F^2$. Taking the ratio, we find that in the Dirac fluid limit $\mu \ll k_B T$,
\begin{align}
    \frac{\sigma_z}{\sigma_F} &\sim \frac{n_+}{n_-} \frac{W/(n_- v_F)^2}{k_B T /v_F^2} \frac{\gamma_\text{imp} + \nu/L^2}{\gamma_\text{ee}} 
    \nonumber
    \\
    &\sim \frac{(k_B T)^2}{\mu^2} \left[\frac{l_\text{ee}^2}{L^2} + \frac{\gamma_\text{imp}}{\gamma_\text{ee}}\right],
\end{align}
where we evaluate $n_+, n_-, W$ using the Fermi-Dirac distribution \cite{Narozhny2019} and use $\nu \sim l_\text{ee}^2 \gamma_\text{ee}$, where the electron-electron scattering length $l_\text{ee} \sim v_F/\gamma_\text{ee}$.

\end{document}